# The MUSE second-generation VLT instrument

Bacon R.[1], Accardo M.[4], Adjali L.[1], Anwand H.[5], Bauer S.[2], Biswas I.[2], Blaizot J.[1], Boudon D.[1], Brau-Nogué S.[6], Brinchmann J.[7], Caillier P.[1], Capoani L.[1], Carollo C.M.[3], Contini T.[6], Couderc P.[6], Daguisé E.[1], Deiries S.[4], Delabre B.[4], Dreizler S.[5], Dubois J.[1], Dupieux M.[6], Dupuy C.[4], Emsellem E.[1,4], Fechner T.[2], Fleischmann A.[5], François M.[1], Gallou G.[6], Gharsa T.[6], Glindemann A.[4], Gojak D.[4], Guiderdoni B.[1], Hansali G.[1], Hahn T.[2], Jarno A.[1], Kelz A.[2], Koehler C.[5], Kosmalski J.[1], Laurent F.[1], Le Floch M.[6], Lilly S.J.[3], Lizon J.-L.[4], Loupias M.[1], Manescau A.[4], Monstein C.[3], Nicklas H.[5], Olaya J-C[2], Parès L.[6], Pasquini L.[4], Pécontal-Rousset A.[1], Pello R.[6], Petit C.[1], Popow E.[2], Reiss R.[4], Remillieux A.[1], Renault E.[1], Roth M.[2], Rupprecht G.[4], Serre D.[7], Schaye J.[7], Soucail G.[6], Steinmetz M.[2], Streicher O.[2], Stuik R.[7], Valentin H.[6], Vernet J.[4], Weilbacher P.[2], Wisotzki L.[2], Yerle N.[6]

(1) Centre de Recherche Astrophysique de Lyon (CRAL, CNRS/University Claude-Bernard Lyon I) 9 avenue Charles André, 69230 Saint-genis-Laval, France
(2) Astrophysikalisches Institut Potsdam, An der Sternwarte 16, 14482 Potsdam, Germany
(3) Institute for Astronomy at ETH, Wolfgang-Pauli-Strasse 27, 8093 Zürich, Switzerland
(4) European Southern Observatory, Karl-Schwarzschild-Str. 2, 85748 Garching, Germany
(5) Institut für Astrophysik (Georg-August University of Göttingen), Friedrich-Hund-Platz 1, 37077 Göttingen, Germany
(6) Laboratoire d'Astrophysique de Toulouse-Tarbes (CNRS/University Paul Sabatier), Observatoire Midi Pyrénées, 14, avenue Edouard Belin, 31400 Toulouse, France
(7) NOVA Leiden Observatory, Leiden University, P.O. Box 9513, 2300 RA Leiden, The Netherlands

Summary: The Multi Unit Spectroscopic Explorer (MUSE) is a second-generation VLT panoramic integral-field spectrograph currently in manufacturing, assembly and integration phase. MUSE has a field of 1x1 arcmin² sampled at 0.2x0.2 arcsec² and is assisted by the VLT ground layer adaptive optics ESO facility using four laser guide stars. The instrument is a large assembly of 24 identical high performance integral field units, each one composed of an advanced image slicer, a spectrograph and a 4kx4k detector. In this paper we review the progress of the manufacturing and report the performance achieved with the first integral field unit.

## 1. Introduction

The Multi Unit Spectroscopic Explorer MUSE is one of the second-generation instruments of the ESO VLT. MUSE is a wide-field optical integral field spectrograph operating in the visible wavelength range with improved spatial resolution.

MUSE has a broad range of astrophysical applications, ranging from the spectroscopic monitoring of solar system's outer planets to very high redshift galaxies. The most challenging scientific and technical application is the study of the progenitors of normal nearby galaxies out to redshifts z>6. These systems are extremely faint and can only be found by their Lyman-alpha emission. MUSE will be able to detect these in large numbers (~15,000) through a set of nested surveys of different area and depth. The deepest survey will require very long integration (80 hrs each field) to reach a limiting flux of $4 \times 10^{-19}$erg.s$^{-1}$.cm$^{-2}$, a factor of 30 times better than what is currently achieved with narrow band imaging.

MUSE is composed of 24 identical modules, each one consisting of an advanced slicer, a spectrograph and a (4k)² detector. A series of fore-optics and splitting and relay optics is in charge of derotating and splitting the square field of view into 24 sub-fields. These are placed on the Nasmyth platform between the VLT Nasmyth focal plane and the 24 IFU modules. AO correction will be performed by the VLT deformable secondary mirror. Four sodium laser guide stars are used, plus a natural star for tip/tilt correction.

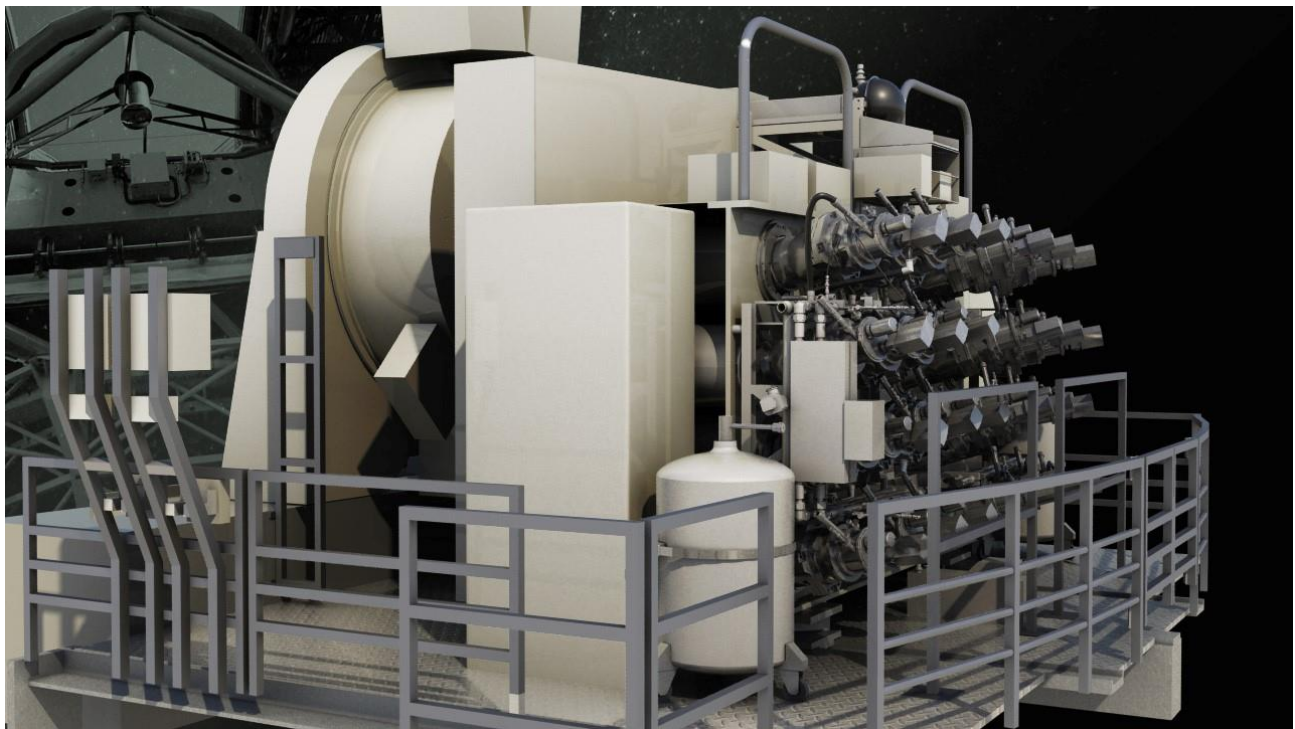
**Figure 1: Artist view of MUSE on the VLT Nasmyth platform**

This unique and ambitious project is supported by seven major European research institutes: the Centre de Recherche Astrophysique de Lyon which is the leading institute, the European Southern Observatory (ESO), the Leiden Observatory (NOVA), the Laboratoire d'Astrophysique de Toulouse-Tarbes, the Institut für Astrophysik of Gottingen, the Institute for Astronomy at ETH and the Astrophysikalisches Institut Potsdam. The MUSE consortium brings together more than a hundred scientists and engineers who cover all the required expertise to build and exploit this unique instrument: optics, mechanics, electronics, cryogenics, signal processing, management, observational and theoretical astrophysics.

The instrument concept and science goals are described in detail in [1]. In this paper we focus on the development status of the project and the first performance validations.

## 2. Technical and managerial challenges

Technical and managerial challenges are the natural counterparts of astrophysical goals. The first two top-level requirements are image quality and throughput. The optical system image quality requirement (1.2 pixel FWHM at the red wavelength) has been set up to maintain the exquisite image quality which will be delivered by the AO facility over the full field of view.

The throughput requirement (up to 40% in the central wavelength range) is essential to detect the very faint galaxies we are looking for. Despite this excellent overall system throughput, observation of these very faint sources will require very long integration (up to 80 hours) and thus high stability is also an important requirement. The instrument must also be very well calibrated. A more detailed description of the requirements is given in [2].

The design, manufacturing and assembly of 24 identical high performance integral field units, each one being an assembly of a complex opto-mechanical system with a large detector and its cryogenic system, is a rather new in the field of optical instrumentation and therefore a big challenge for us. Last but not the least, keeping cost and schedule in this context requires a proper management and quality control [3].

## 3. General status

After 1.5 year of phase A, 4 years of design phase, the project is now in its manufacturing integration and test phase which is expected to last 2.5 years. The past and upcoming milestones are given in the following table.

| Review | Date | Status |
|---|---|---|
| Kick off Phase A | 2002-05 | |
| Conceptual Design | 2004-01 | Passed |
| Optical Preliminary Design | 2006-06 | Passed |
| Preliminary Design | 2007-07 | Passed |
| Optical Final Design | 2007-12 | Passed |
| Final Design | 2009-03 | Passed |
| 1st IFU Manufacturing Item Acceptance | 2010-02 | Passed |
| 2 paths global test | 2011-Q3 | |
| 12 paths global test | 2011-Q4 | |
| Preliminary acceptance in Europe | 2012-Q2 | |
| Preliminary acceptance in Chile | 2012-Q4 | |

With the performance validation of the first IFU (see next section), a major milestone has been achieved recently. The upcoming milestones will be the performance validation of two and then twelve complete optical paths (i.e. from the input telescope focal plane to the detector).

To evaluate the evolution of top-level performances with time we have selected three indicators: the limiting emission line flux (in (cgs unit) for a deep exposure of 80 x 1 hour integration time and for an unresolved galaxy, the wide-field mode (WFM) spatial resolution at 750 nm (FWHM arcsec unit) for typical atmospheric conditions using AO, the Strehl ratio corresponding to the narrow field mode

(NFM) spatial resolution at 650 nm for good atmospheric conditions with AO. To assess the risk we also give the margins with respect to specifications in the two critical areas: throughput in three wavelength ranges (blue-green-red) and instrument image quality. These indicators are given in the following table.

| Perf. Type | Milestone | Indicators | | | Margins | |
|---|---|---|---|---|---|---|
| | | WFM Flux | WFM FWHM | NFM SR | Through. | IQ |
| Dreamed | Phase A | 2.7-4.2 $10^{-19}$ | 0.30-0.50 | 05-10% | | |
| Designed | PDR | 2.6-4.2 $10^{-19}$ | 0.33-0.52 | 04-09% | 14-10-17% | 24-40% |
| | FDR | 2.4-3.9 $10^{-19}$ | 0.30-0.50 | 11% | 17-04-08% | 41-24% |
| Build | PAE | | | | | |
| Real | Comm. | | | | | |

One can note that the indicators are quite stable when going from the Phase A (*dreamed*) to FDR (*designed*) performances and good margins exist. There is even a slight improvement in limiting flux. The long design effort to maintain an excellent throughput while keeping the high image quality has clearly paid off. The next step (*build performances*) has already started with the ongoing measurements of the first integral field unit.

## 4. From design to reality: the first IFU

After such a long period of design and prototype development and testing, the 'first technical light' of the first IFU has been an exciting (and sometime stressing) time.

We first received a Volume Phase Holographic grating (VPHG) from KOSI. This grating has been corrected for the non-conformance in throughput uniformity that was detected on the prototype. It was carefully measured again using our test bench and found to be well within the specifications [4]. The grating was then send to Winligth to be integrated into the spectrograph.

We then received the two major slicer components: the image dissector and the focusing mirror arrays. The alignment and test of this compact and complex system took sometime but finally converge and demonstrate the excellent optical performance of Winligth's realization. However we realize soon that the protected silver coating was progressively degrading. Immediately Winlight started to investigate new coatings and performed a series of environment test to find the suitable coating supplier. The slicer was sent back to Winlight and recoated with a simple temporary metallic coating in order to continue the test sequence.

In the meantime, the first spectrograph from Winlight was received to CRAL almost simultaneously with the first detector and its cryogenic system from ESO. The various test benches realized at CRAL, including part of the control software developed by LATT and the new CCD controller developed by ESO were assembled. Before binding the detector vessel to the spectrograph, we found that the field lens that close the detector vessel, was displaying coating degradation. After inspection and discussion with Winlight, it was concluded that the coating is hygroscopic. The coating malediction was going on!

The first detector had also its non-conformity in the presence of a bright spot that could prevent to make long exposure. A work around was found to be able to conduct the test. Finally all sub-systems were aligned and successfully tested. The system image quality and the detector quantum efficiency were measured to be within the specifications. The detector AR graded coating is not only boosting the quantum efficiency but it also almost suppresses the fringing in the red. A few ghosts due to slicer/spectrograph multiple reflections were measured but they were quite faint and could have been neglected. However an elegant solution was found with an adaptation of the slicer pupil/mask that completely suppress all ghosts. The detail measurements and process of the IFU validation is presented in this conference [5]. One of the technical first light images is shown here.

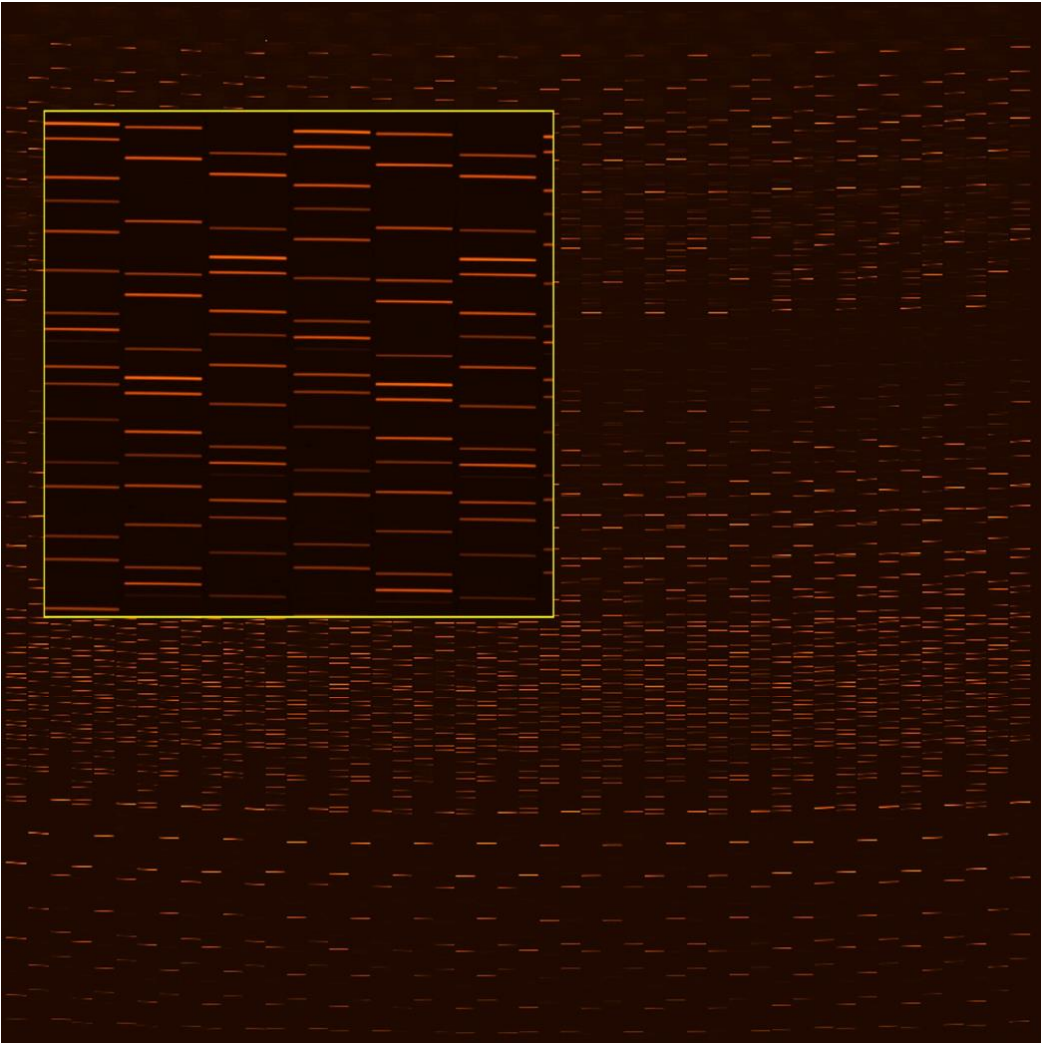

**Figure 2: In this HgCd, Ne and Xe arc exposure of the first IFU one can see 3750 high quality spectra (only 4% of the total number of MUSE spectra)**

Following this first IFU, 23 others are expected. It has then been a significant stress for the Assembly and Integration Test group to face all these non-conformities for the first IFU: finding solution to all these points took a while and the planning slipped consequently. However all the experience gained with this first IFU is essential in view of the series. Actions have been taken to find a good and stable solution to the coating process. The group is now much more confident for the upcoming IFUs. After reviewing the measurements with the various suppliers and ESO, it has been decided to launch the manufacturing of the series. This first IFU will now be refurbished to become fully compliant.

# 5. Status of manufacturing

Even if attention was focused to the first IFU, all other subsytems have well progressed and we have been able to launch in parallel all the manufacturing contracts. An estimate of the progress is also shown in the following table. Note that the 24 detectors have already been delivered by E2V. Performances, as measured by E2V and checked by ESO, are excellent and very uniform between devices.

| Subsystem | Item | Supplier | Progress |
| --- | --- | --- | --- |
| Calibration Unit | Lens | Lichtenknecker Optics | 80% |
| Fore-Optics | Derotator | Winlight | 95% |
| | Mirrors | Hellma | 90% |
| | Lenses | Winlight | 5% |
| | Shutter | Bonn University | 70% |
| | Beam structure | MGS | 70% |
| | Mechanics | Bassanetti | 100% |
| Splitting and Relay Optics | Field Splitter Unit | SEOP | 80% |
| | Relay Optics Mirrors (48x) | SEOP | 5% |
| | Singlet Lenses (18x) | POG | 5% |
| | Doublet Lenses (48x) | Winlight | 5% |
| Instrument Main Structure | Main Structure | Streicher | 85% |
| Integral Field Units | Image Slicers (24x) | Winlight | 30% |
| | Spectrographs (24x) | Winlight | 30% |
| | VPH Gratings (24x) | KOSI | 50% |
| | Detectors (24x) | E2V | 100% |

The cryogenic system is also quite advanced and shall be delivered soon by ESO to CRAL facility. It has already been tested on 12 detector vessels.

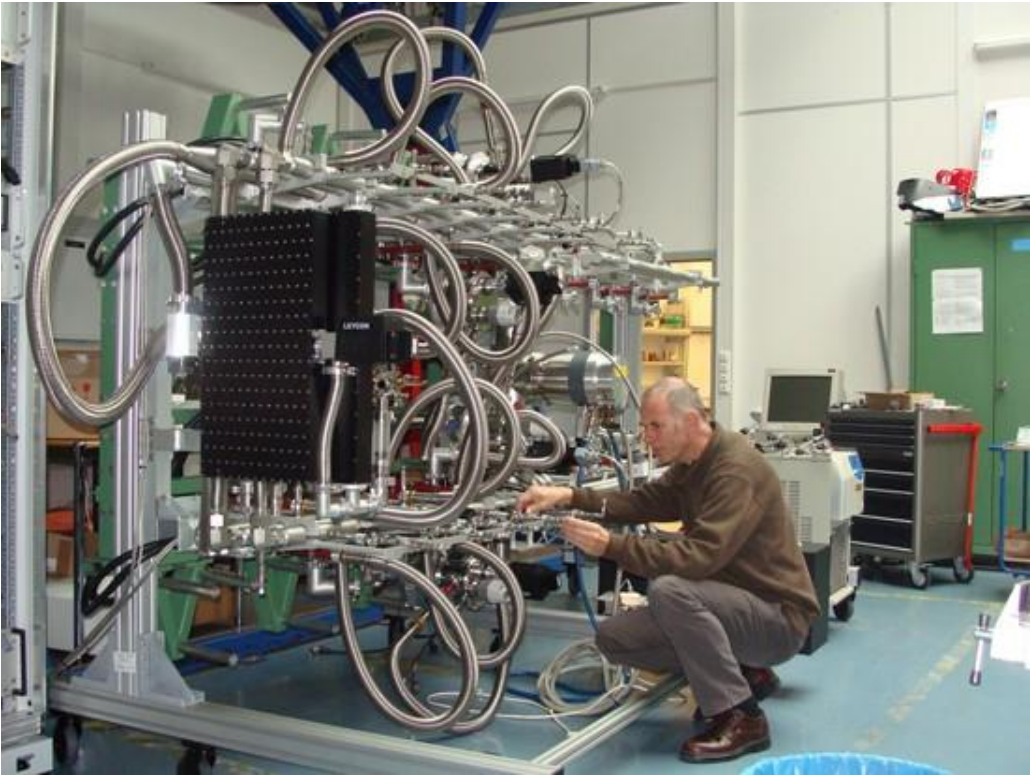
**Figure 3: MUSE cryogenic system in test at ESO/Munich.**

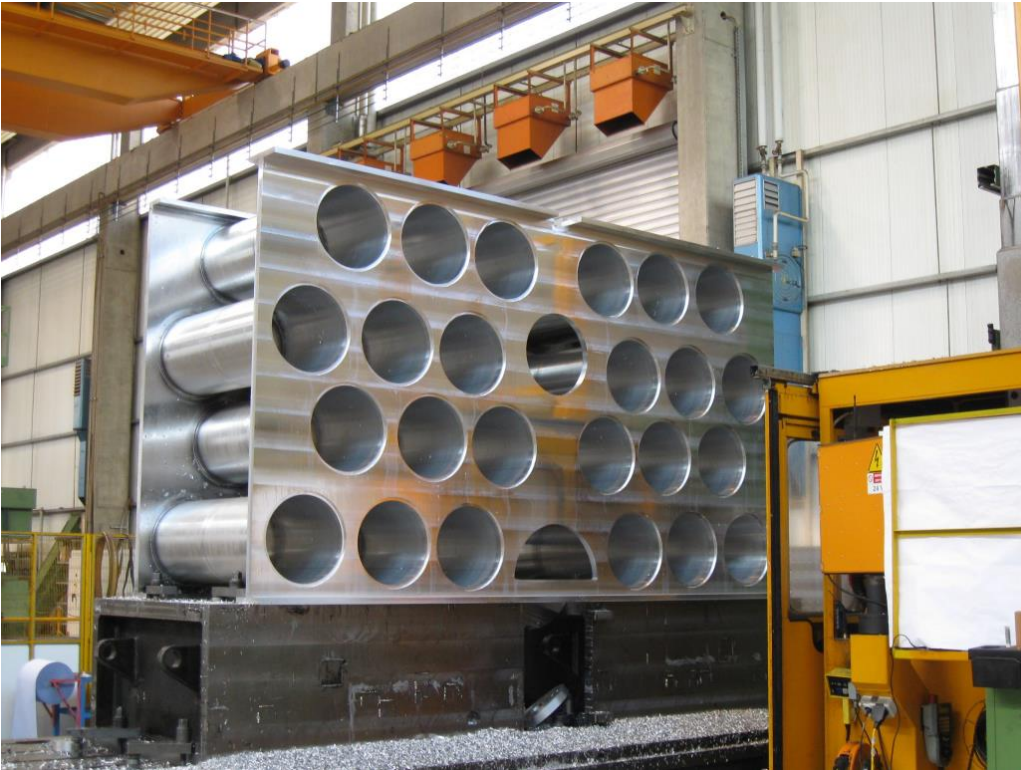
**Figure 4: The large structure with its 24 cylindrical IFU supports in manufacturing stage at Streicher premise.**

## 6. Software development

A first prototype of the instrument control software has been produced for the IFU AIT. In parallel the study of the non-standard software components (the slow guiding system and the fast reconstruction image display) has been finalized. The first system uses a specific detector to perform low order guiding on faint stars while the second is able to compute and display on-the fly a reconstructed white light image using a short exposure of the field of view. The CCD acquisition is performed by the New Generation Controller developed at ESO.

The second major piece is the data reduction software. Preliminary releases have been distributed and tested against simulated data given by the instrument numerical model [3] and on real data obtained with the first IFU. An important feature of the data reduction software is a proper handling of noise propagation.

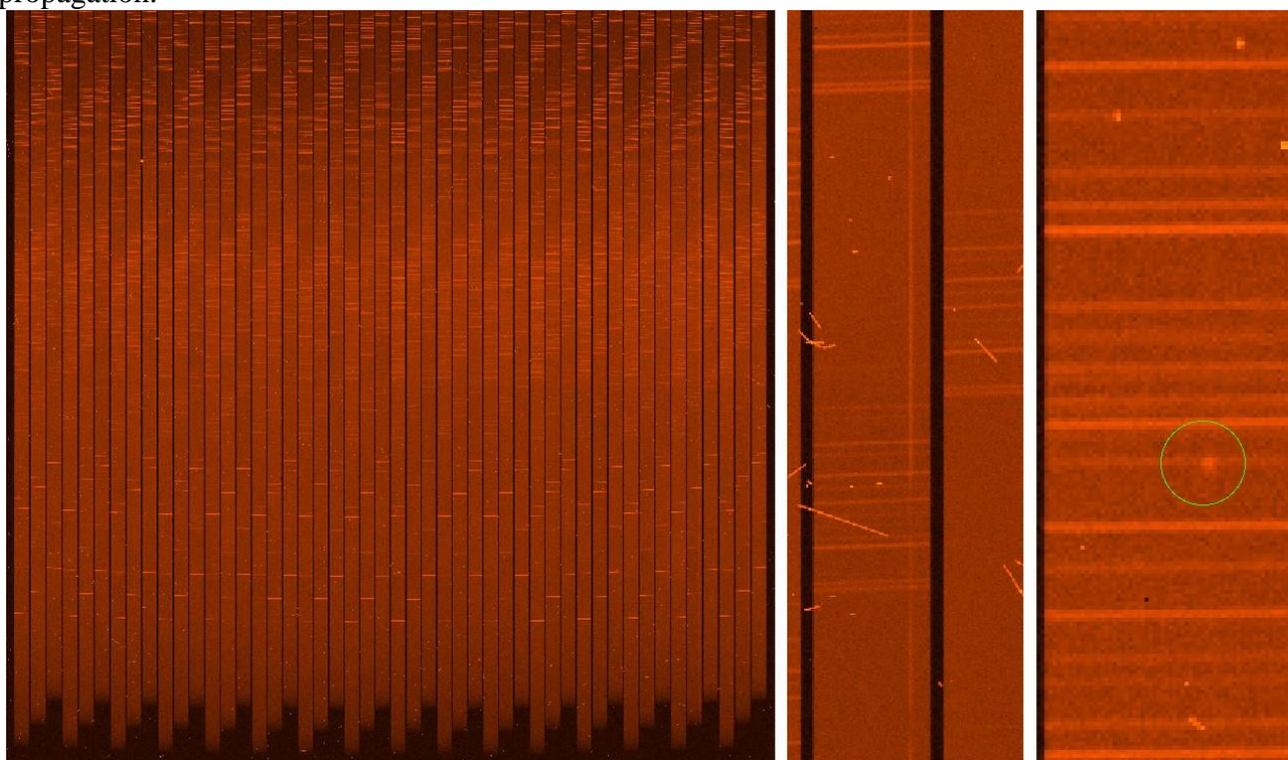

**Figure 5: Instrument numerical model simulation of a raw CCD (bias subtracted) frame corresponding to a single deep field one-hour exposure. From left to right: one full CCD raw exposure, a zoom centered on a faint continuum galaxy and on an emission line object.**

## 7. Conclusions

The design, manufacturing, alignment and test of the first IFU have been successfully performed. Although the system is not fully compliant, solutions have been found to the non-conformities and the manufacturing of the series has been launched. The realization of such a high performance system for a unit cost (detector and its cryogenic system included) of less than 200 k€ is already a major achievement.

The next major milestone is the preliminary acceptance in Europe foreseen for the first quarter of 2012. We expect to get all IFUs components delivered to Lyon by early 2011 and the other hardware (main structure, fore-optics, field and splitter optics) and software a few months later. Then the assembly of all parts ("the grand ballet") will start followed by end-to-end system tests. This impressive machine shall then be ready to be scrutinized by ESO and after successful validation, sent to Paranal for first light at the VLT.